\documentclass{article}

\usepackage{PRIMEarxiv}

\usepackage[utf8]{inputenc} 
\usepackage[T1]{fontenc}    
\usepackage[hidelinks]{hyperref}       
\usepackage{url}            
\usepackage{booktabs}       
\usepackage{amsfonts}       
\usepackage{nicefrac}       
\usepackage{microtype}      
\usepackage{lipsum}
\usepackage{fancyhdr}       
\usepackage{graphicx}       
\usepackage{subcaption}
\usepackage{tabularx}
\usepackage[numbers,sort&compress]{natbib}
\usepackage{amsmath}
\graphicspath{{media/}}     

\title{NOx emissions trends in hydrogen lean premixed flamelets at high strain rate
}

\author{
  A. Porcarelli$^{*}$, B. Kruljevic and I. Langella \\
  TU Delft, Faculty of Aerospace Engineering, Flight Performance \& Propulsion group  \\
  Kluyverweg 1, 2629 HS Delft, Netherlands \\
  \texttt{*\href{mailto:a.porcarelli@tudelft.nl}{a.porcarelli@tudelft.nl}} \\
}

\begin{document}
\maketitle

\begin{abstract}
NO$_{\rm x}$ formation in lean premixed and highly-strained pure hydrogen-air flamelets is investigated numerically. Lean conditions are established at an equivalence ratio of 0.7. Detailed-chemistry, one-dimensional simulations are performed on a reactants-to-products counter-flow configuration with an applied strain rate ranging from $a=100 \, {\rm s}^{-1}$ to $a=10000 \, {\rm s}^{-1}$ and the \texttt{GRI3.0} mechanism. Following a similar setup, two-dimensional direct numerical simulations are also conducted for representative strain rates of $2000 \, {\rm s}^{-1}$ and $5000 \, {\rm s}^{-1}$. Both solutions show a decreasing NO$_{\rm x}$ trend as the applied strain rate is increased. This decreasing emission outcome is highlighted for the first time in this study for lean pure-hydrogen flamelets. A deep analysis of the 2D solution underlines that there is no production of NO$_{\rm x}$ in the second dimension, thus proving that the emission trend is not a result of a setup preconditioning, but is instead a direct physical effect of stretch on the flame. Furthermore, a detailed analysis of the NO$_{\rm x}$ formation pathways at $a=2000 \, {\rm s}^{-1}$ and $a=5000 \, {\rm s}^{-1}$ is performed. Thermal NO$_{\rm x}$ and NNH pathways are shown to both contribute significantly to the total NO$_{\rm x}$ production. While the NNH route contribution is roughly constant at different strain rates, a significant decrease is observed along the thermal NO$_{\rm x}$ route. Overall, results show that lean and highly-strained hydrogen flames experience a significant decrease of NO$_{\rm x}$. This property is discussed and analysed in the paper.
\end{abstract}


\section*{Introduction}
\begin{figure}[t]
	\centering
	\begin{subfigure}{0.35\textwidth}
         \centering
         \includegraphics[width=\textwidth]{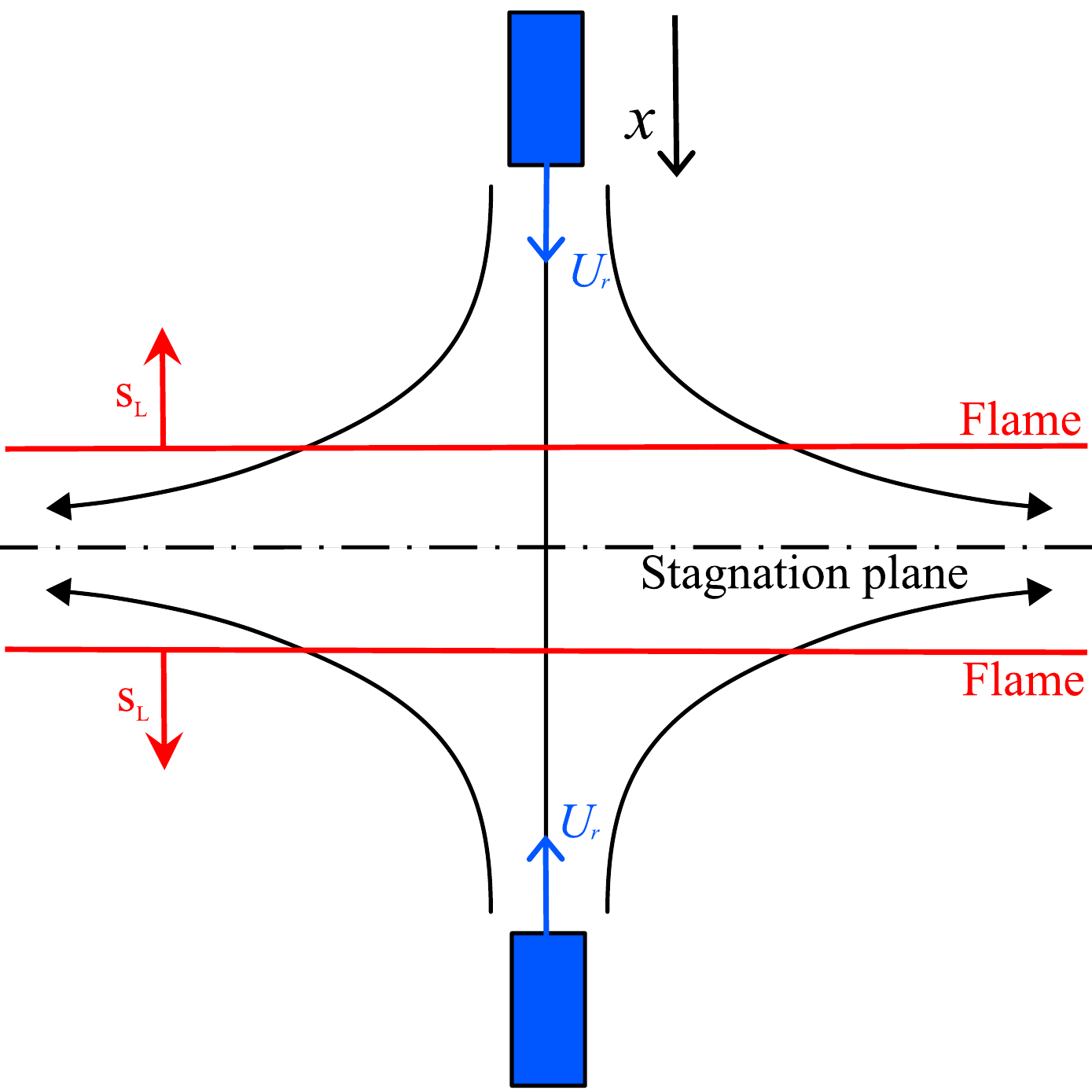}
         \caption{}
         \label{fig:counterflow_twin}
     \end{subfigure}
     \begin{subfigure}{0.35\textwidth}
         \centering
         \includegraphics[width=\textwidth]{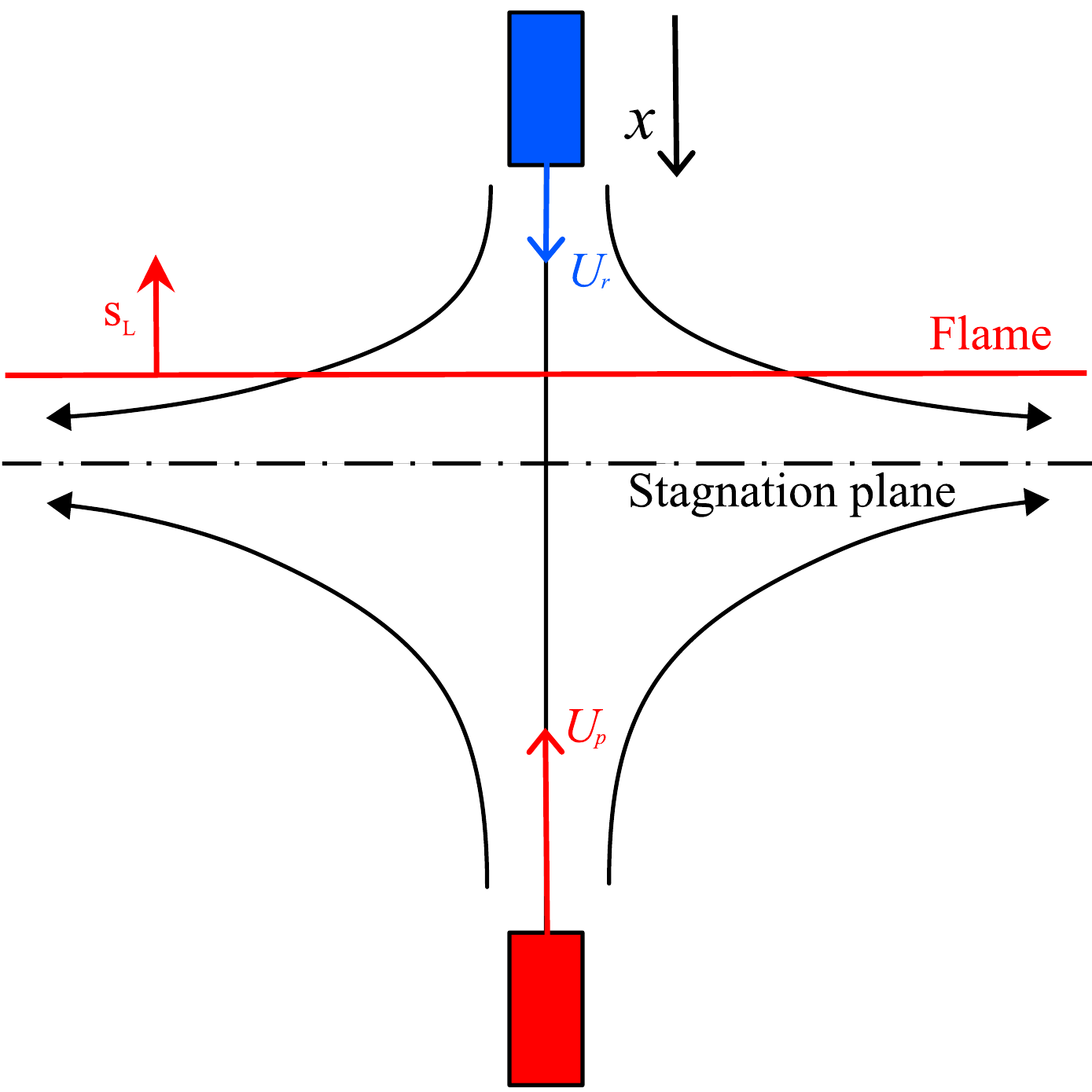}
         \caption{}
         \label{fig:counterflow_products}
     \end{subfigure}
	\caption{Sketch of the reactants-to-reactsants (a) and reactants-to-products (b) counter-flow premixed strained flame configurations.}
	\label{fig:counterflow_config}
\end{figure}

The yearly increasing energy demand is currently substantially met by fossil fuels, with consequent emissions of CO$_2$ and other pollutants. Hydrogen has been identified as one of the possible alternatives to satisfy this demand and reduce green-house emissions simultaneously. In fact, hydrogen is carbon-free, and can be produced with electrolysis using renewable energy \cite{turner2004sustainable,jensen2007hydrogen}. In the last decades, research efforts have focused on the possibility to burn hydrogen in lean premixed conditions, where the lower adiabatic flame temperature allows to decrease NO$_{\rm x}$ formation via the thermal route. 
Furthermore, hydrogen's strong reactivity and high lower-heating-value allows to reach ultra-lean conditions without approaching lean blow-off. Earlier studies have discussed the influence of hydrogen addition on lean blow-off, showing that a minimal hydrogen enrichment can consistently decrease the blow-off equivalence ratio and thus broaden the possible burning regimes \cite{jackson2003influence, cho2009improvement, barbosa2007control,lipatnikov2005molecular}.  However, many technological challenges are involved with controlling hydrogen combustion in lean turbulent conditions. In fact, flashback and uncontrolled flame propagation that are typical of this regime \cite{huang2009dynamics,ducruix2003combustion} are amplified by the combination of hydrogen's high reactivity and so its higher flame speed \cite{speth2009using}, and its ability to auto-ignite \cite{hu2019large,johannessen2015experimental} and quickly diffuse \cite{dinesh2016high}. \\ 
One interesting property of hydrogen flames that has not been fully understood yet is its behaviour under strain. Previous studies have highlighted the peculiar performance of hydrogen-enriched laminar flames with varying strain rates, suggesting that high strain regimes can potentially be exploited for hydrogen combustion practical applications. Hydrogen addition has been shown to delay the extinction strain rate \cite{jackson2003influence,cho2009improvement}, proving that hydrogen is able to sustain very high strain rates. It has also been shown for syngas fuels that hydrogen percentage influences the lean flame response with strain in terms of flame temperature and NO$_{\rm x}$ emissions \cite{ning2017effects}, and flame temperature and consumption speed \cite{speth2009using}. In particular, in lean conditions the consumption speed is shown to increase with strain, and this trend is opposite to the one observed for hydrocarbon flames \cite{liang2017extrapolation}. Furthermore, the hydrogen-enhanced differential diffusion effects and the strain sensitivity have been proved to be interdependent and to have a combined influence on the flame response \cite{abbasi2019differential,vance2022quantifying}.
Even more interestingly, while the mass burning rate decreases with strain for pure methane flames, a certain amount of hydrogen addition inverts this trend, particularly in lean conditions \cite{van2016state}. Similar considerations hold for the heat release rate, which is found to increase with strain in lean hydrogen enriched flames \cite{marzouk2005combined}. However, the effect of these distinctive hydrogen burning features at high strain rates on NO$_{\rm x}$ emission for purely hydrogen lean laminar flames is still an open question. \\
In this study we numerically investigate counter-flow premixed hydrogen flames to shed light on their behaviour under intensive level of strain, in particular the effect on NO$_{\rm x}$ emissions. This is, to best of the authors' knowledge, the first time that such an investigation is conducted. The increase in heat release rate with strain discussed in the previous paragraph would suggest a corresponding increase of temperature and so of NO$_{\rm x}$ emissions. Counter-intuitively, this study highlights 
that NO$_{\rm x}$ emissions do not increase with strain, and conversely show a decreasing trend in lean conditions, particularly at very high strain rates. Nevertheless, despite the discussed advantages and potentialities, additional unexplored control challenges are potentially introduced at high strain regimes, such as complex vortex dynamics, turbulence-flame interaction and local flame extinctions. Further modelling challenges are associated with the prediction of these phenomena distinctive of the regime. \\
As reported in Figure \ref{fig:counterflow_config}, there are two premixed strained flamelet configurations documented in literature.
The first one is represented by a back-to-back or reactants-to-reactants counter-flow configuration, where two flames are stabilized symmetrically with respect to the flow stagnation plane (Figure \ref{fig:counterflow_twin}). Several studies are available where the effects of hydrogen enrichment and strain on lean blow-off, extinction strain, mass burning rate, and NO$_{\rm x}$ emissions are investigated with this setup \cite{cho2009improvement,jackson2003influence,van2016state,ning2017effects}. Specifically on NO$_{\rm x}$ emissions trends, a technical report from \citet{xie2013analysis} underlines a decreasing trend with strain with this configuration in rich pure-hydrogen flames, attributing to the NNH pathway the predominant contribution to NO$_{\rm x}$ formation at high strain levels. 
However, the limitation of this configuration is represented by the proximity of the flames to the stagnation plane, particularly at the high strain rates achievable by purely hydrogen flames. In these conditions, in fact, the combustion has no space to complete, as it is clearly visible in previous studies. Minor species are shown not be fully burnt within the stagnation plane in Figure 5 and 6 of Jackson et al. \cite{jackson2003influence}. Similarly, the reaction rates of NO$_{\rm x}$ pathways  achieve zero within the stagnation plane for lower strain rates in Ning et al. \cite{ning2017effects} (see Figures 6-8 of their work), but not at higher strain rates. The same happens to the water rate of production in \citet{xie2013analysis}, Figure 2. Hence, 
this configuration appears less suitable for investigations on emissions, and ultimately not practical for the development of flamelet databases for turbulent LES simulations. \\
The second  premixed strained flamelet configuration is represented by a reactants-to-products counter-flow, where a single flame stabilizes on the reactants side of the domain (Figure \ref{fig:counterflow_products}). On the one hand, the single flame setup allows the combustion reactions to complete even at very high strain rates, as the fuel and the radicals have space to be burnt completely. This evidence can be found for instance in \citet{marzouk2005combined}, Figure 4-6. This consideration suggests that the configuration is more appropriate for emission analyses in highly-strained premixed flamelets. On the other hand, the presence of complete combustion hot products on one of the boundaries can precondition the problem, particularly considering the products temperature. Despite this, the reactants-to-products configuration has been employed  for physical analysis of strained syngas and hydrogen-enriched flames \cite{speth2009using,marzouk2005combined}, and widely discussed for methane flamelets \cite{libby1982structure,darabiha1986effect}, particularly in the application of LES-FGM models with strained flamelets \cite{kolla2010strained,langella2016unstrained}. Therefore, the reactants-to-products configuration is employed in the present study. \\
The purpose of the present work is to investigate lean hydrogen strained flamelets, with a particular focus on NO$_{\rm x}$ emissions trends.  
Detailed-chemistry one-dimensional and two-dimensional DNS analyses are performed to achieve a deeper understanding of the employed counter-flow strained flamelet configuration. It will be shown for the first time that NO$_{\rm x}$ emissions of purely-hydrogen lean laminar flames display a decreasing trend with strain. A detailed chemical analysis of NO$_{\rm x}$ formation pathways is performed to support this evidence, further showing that thermal NO$_{\rm x}$ mechanism is predominant. This physical behaviour of laminar flames  is the first step to shed light on potential features of novel combustor systems, where NO$_{\rm x}$ emissions are controlled by stabilising the flame against intensive strain. \\
This paper is organised as follows. The flamelet equations along with the modelling choices performed for the 1D and 2D cases are introduced first, followed by an overview of the 1D and 2D numerical setups. Emission results are discussed next. 
Finally, relevant conclusions are drawn on the influence of strain on NO$_{\rm x}$ emissions.

\section*{Model}
Both one-dimensional and two-dimensional simulations are performed. For the two-dimensional simulations, the reactingFoam solver in OpenFOAM \cite{weller1998tensorial} is employed. The reacting Navier Stokes equations~\cite{poinsot2005theoretical} are solved for mass, momentum, absolute enthalpy 
and $N$ species with detailed chemistry. The equation for the generic species $k$ is

\begin{equation}
       \label{eq:species}
       \frac{\partial(\rho Y_k)}{\partial t}+\frac{\partial (\rho u_i Y_k)}{\partial x_i} = -\frac{\partial (\rho {V}_{k,i} Y_k)}{\partial x_i} + W_k \dot{\omega}_k,
\end{equation}
where subscripts $i$ denotes direction $i$, 
$W_k$ is the molar mass, $\rho$ the mixture density, $V_{k,i}$ the diffusion velocity vector, and $\dot{\omega}_k$ the molar rate of production of species $k$. 
A low Mach number approximation is used in this study. Radiation, body forces, and viscous dissipation effects are neglected. The Dufour effect on the heat flux is also neglected. \\
The ideal gas law and the caloric equation of state are used as thermodynamic model, where the species heat capacities are obtained using the JANAF polynomials. Only laminar conditions are considered in this study, while the effect of the turbulence-chemistry interaction will be investigated in a future study. 
Detailed kinetic data of reactions from \texttt{GRI3.0} mechanism \cite{GRI30} are used to obtain consumption and production rate of species. In order to speed up the simulations, the \texttt{TDACChemistryModel} is adopted, consisting of the combination of the \textit{in situ} adaptive tabulation (ISAT) algorithm with the dynamic adaptive chemistry (DAC) reduction scheme \cite{Contino2011}. 
A mixture-averaged diffusion model is used \cite{poinsot2005theoretical} to account for the low Lewis number of the hydrogen fuel as follows:
\begin{subequations}
    \begin{equation}
        \label{eq:fick}
        V_{k,i} = - \frac{D_k^M}{X_i} \frac{\partial X_k}{\partial x_i} \approx \frac{D_k^M}{Y_i} \frac{\partial Y_k}{\partial x_i}
    \end{equation}
    \begin{equation}
        \label{eq:diffcoeffs}
        D_k^M = \frac{1-Y_k}{\sum_{l,k\neq l}^{N} {X_l}/{D_{kl}}}
    \end{equation}
    \label{eq:of_diff}
\end{subequations}
where the validity of the approximation performed in Eq.~\eqref{eq:fick} is verified in post-processing. The binary diffusion coefficients $D_{kl}$ for the species involved in the reactions are found with the Chapman-Enskog correlation \cite{chapman1990mathematical,Novaresio2012}. \\
 One-dimensional simulations are  run with CHEM1D \cite{chem1d}. In one dimension and for a flat reactants-to-products counter-flow flame, the set of conservation equations solved in CHEM1D is re-arranged as follows \cite{ramaekers2011development}:
\begin{subequations}
    \begin{equation}
        \label{eq:mass1D}
         \frac{\partial\rho}{\partial t} + \frac{\partial (\rho u_x)}{\partial x} = -\rho K
    \end{equation}
    \begin{equation}
        \label{eq:species1D}
        \frac{\partial(\rho Y_k)}{\partial t}+\frac{\partial(\rho u_x Y_k)}{\partial x} =  -\frac{\partial (\rho V_{x,k} Y_k)}{\partial x} + W_k\dot{\omega}_k -\rho K Y_k
    \end{equation}
    \begin{equation}
        \label{eq:momentum1D}
        \frac{\partial\rho K}{\partial t}+\frac{\partial\rho u_x K}{\partial x} = \frac{\partial}{\partial x}\left(\mu\frac{\partial K}{\partial x}\right)+ \rho_{\rm p}a^2 -\rho K^2
    \end{equation}
    \begin{equation}
        \label{eq:enthalpy1D}
         \frac{\partial\rho h}{\partial t}+\frac{\partial\rho u_x h}{\partial x} = \frac{\partial q}{\partial x} -\rho K h
    \end{equation}
    \label{eq:governing1D}
\end{subequations}
where the density of the products mixture $\rho_{\rm p}$, the applied strain rate $a$, the local stretch rate $K$, along with Newton's law for viscous stresses have been introduced. The applied strain rate is a setup parameter and is defined as the velocity gradient at the products boundary:
\begin{equation}
    \label{eq:a}
    a = -\left(\frac{du_x}{dx}\right)_{\rm p}.
\end{equation}
The influence of the $y$-component of the flow on the transport of scalars is taken into account with the introduction of the local stretch rate \cite{ramaekers2011development}:
\begin{equation}
    \label{eq:K}
    K(x) = \frac{\partial u_y}{\partial y}.
\end{equation}
As a consequence of the continuity equation, the relation between the two parameters reads $K(x\rightarrow\infty) = a$. Similarly to two-dimensional simulations, \texttt{GRI3.0} is used a chemical mechanism, along with a mixture-averaged diffusion model (Eq.~\eqref{eq:of_diff}), while the binary diffusion coefficients in CHEM1D are computed using molecular potentials and are tabulated as a function of temperature in polynomial form \cite{somers1994simulation}. 

\section*{Computational setup} \label{sec:setup}
The combustion of hydrogen is evaluated at atmospheric conditions. Temperature and species boundary conditions at the reactants and products boundary are prescribed as follows for both 1D and 2D cases. An equivalence ratio of $\phi=0.7$ is imposed for both streams. The reactants temperature is $T_{\rm r}=300 \, {\rm K}$, while the products temperature is set to the adiabatic flame temperature of an unstrained hydrogen flamelet with the same equivalence ratio computed with CHEM1D, $T_{\rm p}=2021 \, {\rm K}$. Mass fractions at the products boundary are imposed from complete combustion.
A summary of the temperature and species boundary conditions for both 1D and 2D simulations are reported in Table \ref{tab:species&T}. \\
\begin{table}[b]
	\centering
	\caption{Temperature and species boundary conditions for both 1D and 2D simulations.}
	\vspace{8pt}
	\begin{tabular}{c c c}
		\hline\noalign{\smallskip}
		Quantity & Left boundary (reactants) & Right boundary (products)  \\
		\noalign{\smallskip}\hline\noalign{\smallskip}
		T [K] & 300 & 2021 \\
		$Y_{\rm H_2}$ [-] & 0.02 & 0 \\
		$Y_{\rm O_2}$ [-] & 0.228 & 0.068 \\
		$Y_{\rm H_2O}$ [-] & 0 & 0.18 \\
		$Y_{\rm N_2}$ [-] & 0.752 & 0.752 \\
		\noalign{\smallskip}\hline
	\end{tabular}
	\label{tab:species&T}
\end{table}

\subsection*{1D setup}
For the one-dimensional simulations, a wide computational domain of $L_{\rm 1D}=20$ cm is chosen. CHEM1D uses an adaptive mesh algorithm, thus ensuring that a proper mesh refinement in the flame region. The adaptive grid accounts for 200 points in total. An exponential differential scheme is used for the spatial discretization, and a second-order time integration of the differential equations is performed by the stationary solver. The time step is adjusted automatically by the numerical tool to achieve convergence, and ranges between $10^{-6} \, {\rm s}$ and $10^{-8} \, {\rm s}$. 
The only input parameter for the velocity field is represented by the applied strain rate $a$, defined in Eq.~\eqref{eq:a}. According to the applied strain rate definition, the higher is $a$, the higher will the boundary velocities be and the stretch rate experienced by the flame. 
A broad range of applied strain rates is investigated for 1D simulations, from $100 \, {\rm s}^{-1}$ up to $10000 \, {\rm s}^{-1}$ to eventually observe the occurrence of flame extinction. The simulations wall clock time ranges from a few minutes for very high strain rates up to a few hours for lower strain rates.

\subsection*{2D setup}
\begin{figure}[t]
	\centering
	\includegraphics[width=\textwidth]{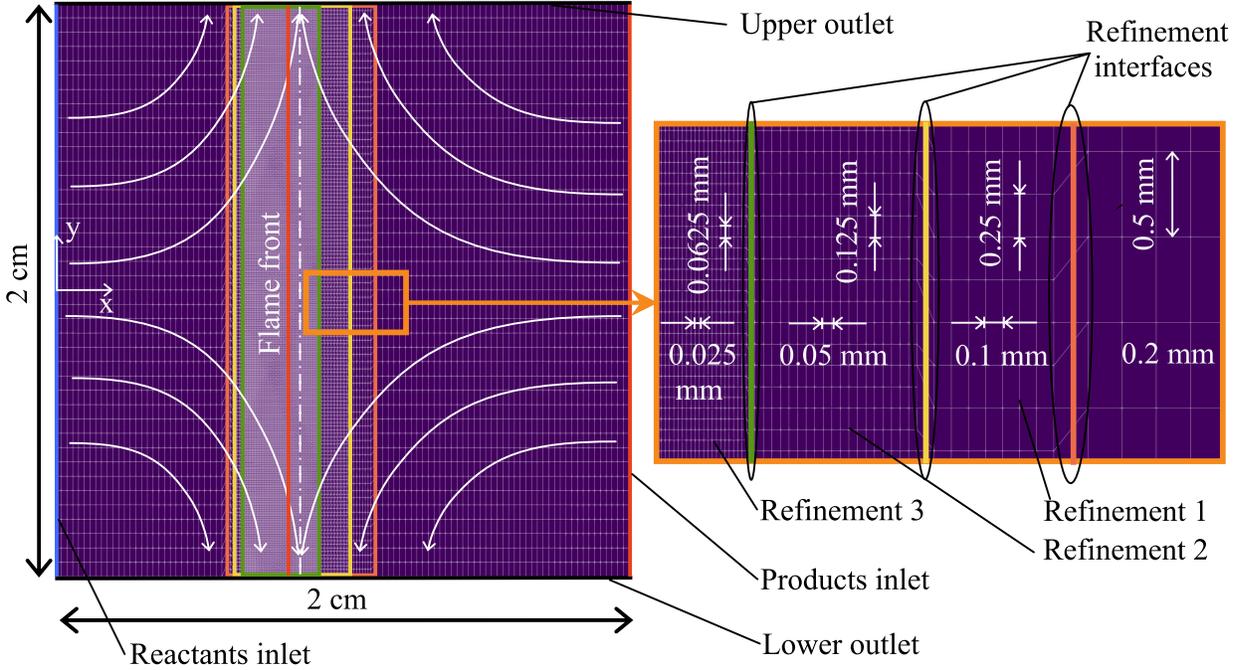}
	\caption{Sketch of the two-dimensional setup with mesh refinement for $a=5000 \, {\rm s}^{-1}$.}
	\label{fig:mesh}
\end{figure}

The PISO time integration solver \cite{issa1986solution}, with an implicit second-order Euler-backward discretization scheme, is used for the 2D simulations. Similarly, a second-order central scheme is used for the convective term of all resolved quantities. 
A variable time step is used ensuring a maximum Courant number of 0.5 for all domain. The simulations are run for a physical time of $t=0.5 \, {\rm s}$ for both the applied strain rates investigated, corresponding to more than 100 times the maximum estimated flow-through time inside the domain. 
A typical computation required 225 and 895 CPU-hours respectively for lower and higher strain rate cases. \\
The higher computational cost of two-dimensional DNS simulations led to the choice of decreasing the dimension of the domain 
to $L_{\rm 2D}=2$ cm in both the horizontal and the vertical direction. Nevertheless, this domain  was sufficient to observe a flame position enough far from the boundary for all simulations. The parent mesh is made of 100 uniform quadrilaterals in the x-direction and 40 in the y-direction, resulting into $\Delta x = 0.2$ mm and $\Delta y = 0.5$ mm. Hence, the quads are refined progressively by a factor of 2 for three times in the region in the proximity of the flame. 
An example for $a=5000 \, {\rm s}^{-1}$ is provided in Figure \ref{fig:mesh}. The flame position is determined for every applied strain rate combining preliminary coarse 2D simulations and 1D simulations, and is then double-checked in the refined simulation solution \textit{a posteriori}. After the refinement, the minimum spacing results into $\Delta x_{\rm min} = 0.2/2^3=0.025$ mm, which ensures at least 12 cells are present within the flame thickness. The number of cells enclosed within the flame thickness is similar to that found in previous DNS studies on reacting flows \cite{klein2020evaluation,lee2022dns,berger2020dns}. 
In contrast to CHEM1D, where the applied strain rate $a$ was the only input parameter for the velocity field, a uniform velocity at the two inlets is prescribed in OpenFOAM's setup. Given the extended-domain one-dimensional simulation at a given applied strain rate, the following horizontal velocity boundary conditions are thus applied in the 2D case:
\begin{subequations}
    \label{eq:Ubc}
    \begin{equation}
        \label{eq:Ubcreactnts}
         \left( U_{\rm r,2D} \right)_{a= \rm const}= \left( U_{\rm 1D}(x=-1 \, \rm{cm}) \right)_{a= \rm const},
    \end{equation}
    \begin{equation}
        \label{eq:Ubproducts}
        \left( U_{\rm p,2D} \right)_{a= \rm const}= \left(U_{\rm 1D}(x=1 \, \rm{cm}) \right)_{a= \rm const}.
    \end{equation}
\end{subequations}
Hence, at the upper and lower outlets, a zero gradient boundary condition is prescribed. The discrepancy in the inlet boundary conditions definition precludes the possibility to directly compare 1D and 2D simulations at a given applied strain rate, because the resulting 2D horizontal velocity profile does not perfectly overlap with the one in the 1D solution. However, as it will be further discussed in the Results section, the purpose of two-dimensional simulations is solely to show that the same NO$_{\rm x}$ emission trends with strain are observed with respect to 1D simulations. Therefore, this setup discrepancy is accepted for the objectives of the present study. Furthermore, only two high applied strain rates configurations are investigated in the 2D case, $a=2000 \, {\rm s}^{-1}$ and $a=5000 \, {\rm s}^{-1}$.

\section*{Results} \label{sec:results}

\subsection*{NO$_{\rm x}$ emissions trends}
\begin{figure}[t]
	\centering
	\includegraphics[width=0.8\textwidth]{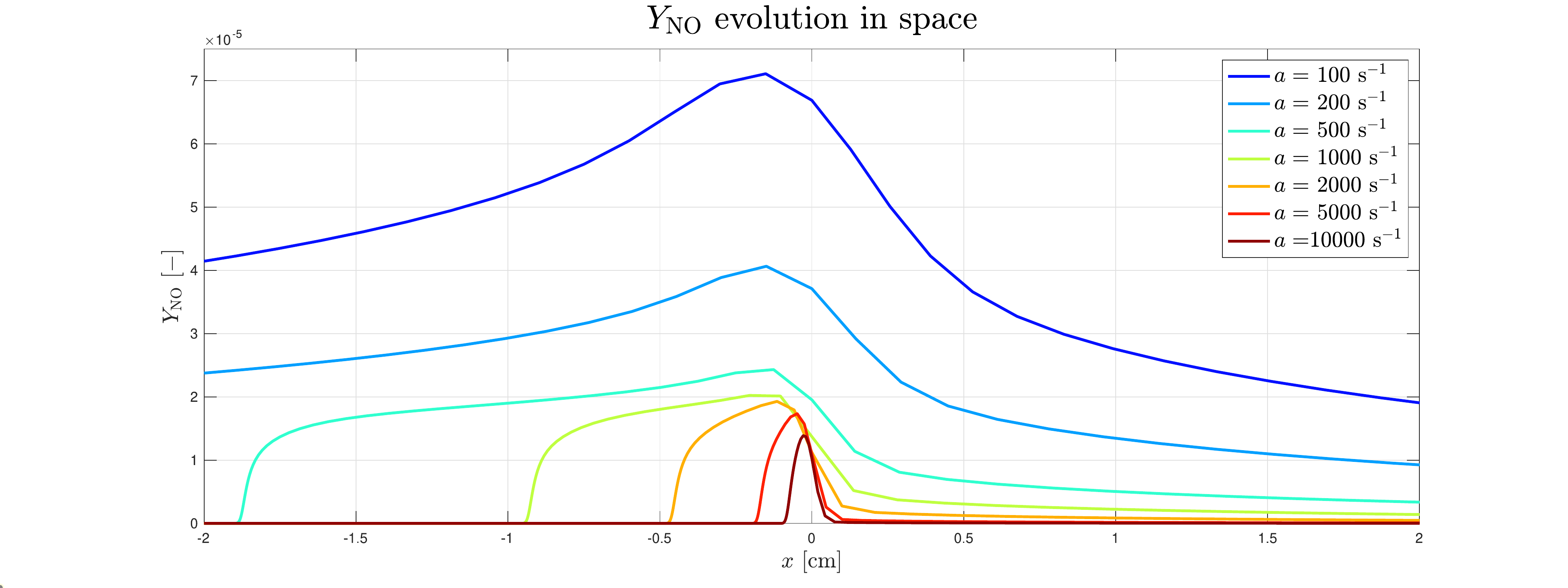}
	\caption{NO mass fraction as a function space for different applied strain rates.}
	\label{fig:NOvsx}
\end{figure}

\begin{figure}[t]
	\begin{subfigure}{0.45\textwidth}
         \centering
         \includegraphics[width=\textwidth]{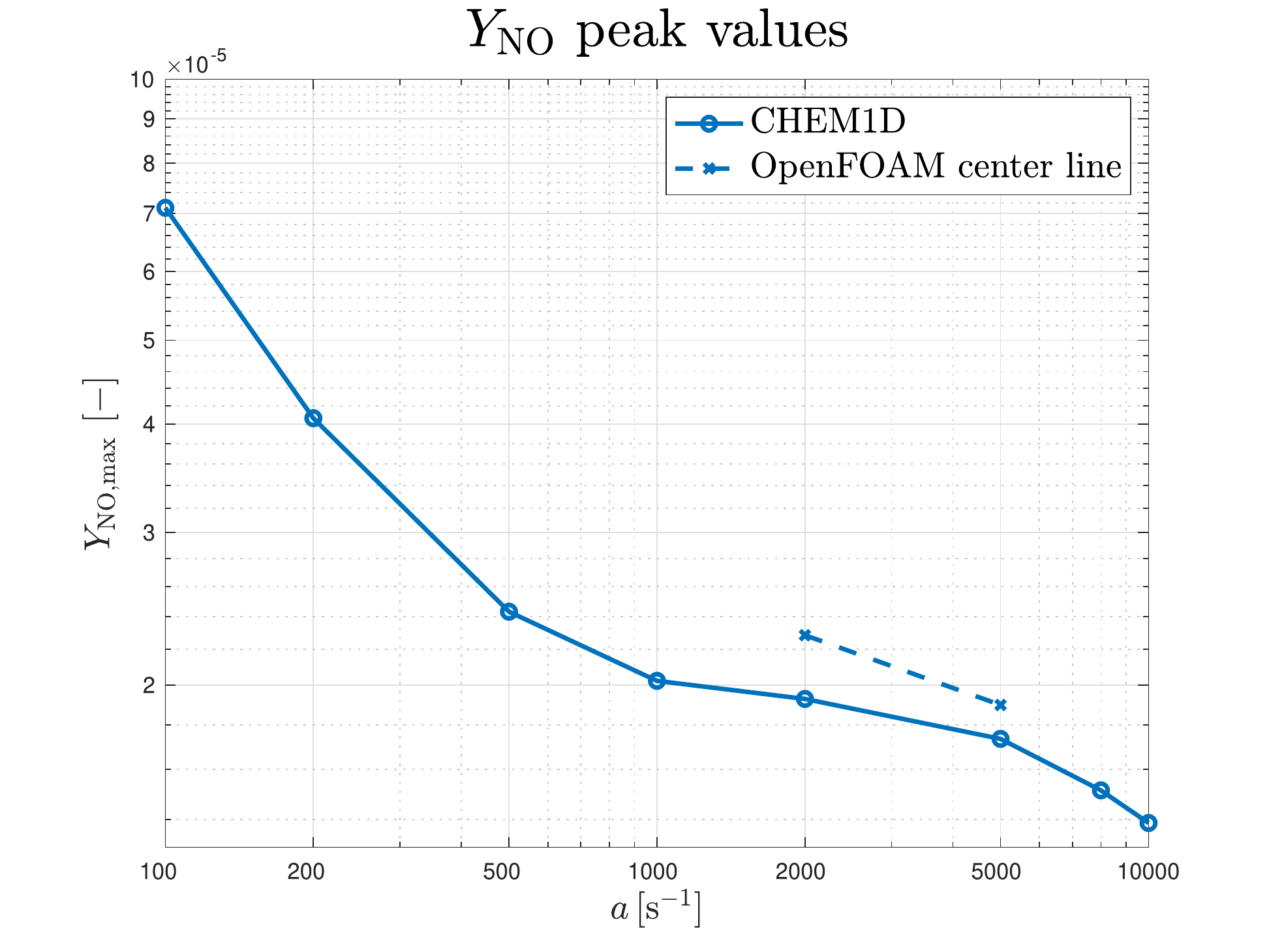}
         \caption{}
         \label{fig:NOpeaks}
     \end{subfigure}
     \begin{subfigure}{0.45\textwidth}
         \includegraphics[width=\textwidth]{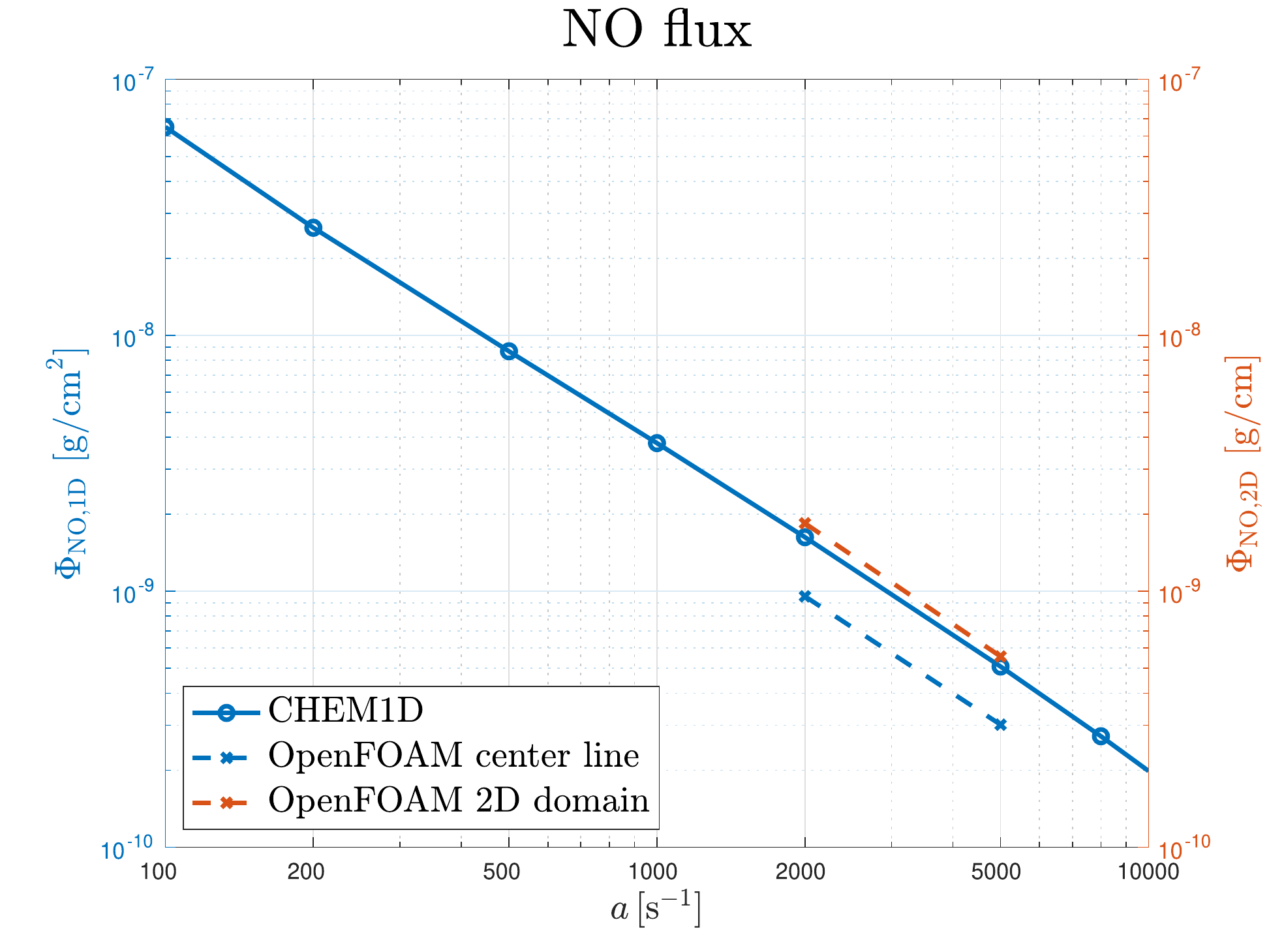}
         \caption{}
         \label{fig:NOfluxes}
     \end{subfigure}
	\caption{Variation of NO mass fraction (a) and NO flux (b) with strain.}
	\label{fig:peaks&fluxes}
\end{figure}

NO$_{\rm x}$ generally refers to nitrogen oxides, including both NO and NO$_2$. However, NO$_2$ mass fraction peaks are shown to be at least three orders of magnitude smaller than the ones of NO for any setup investigated. Therefore, similarly to previous studies \cite{ning2017effects}, NO emissions can alone be considered representative of the overall NO$_{\rm x}$ emissions and are thus examined for the purposes of the present study. The behaviour of $Y_{\rm NO}$ in the longitudinal direction at different applied strain rates predicted by CHEM1D in the proximity of the stagnation plane is shown in Figure \ref{fig:NOvsx}. It is immediately visible that both the peak and the area under the curve decrease consistently as the applied strain is increased. In addition, no extinction is observed at the highest strain rates achieved. As one could expect considering previous studies involving hydrogen-enriched fuels \cite{jackson2003influence, cho2009improvement}, this evidence confirms that pure hydrogen flames can potentially sustain very high strain levels. In Figure \ref{fig:NOpeaks} the decrease of the peaks of $Y_{\rm NO}$ in one-dimensional and the center line of two-dimensional simulations is further highlighted. The discrepancies in the exact values of $Y_{\rm NO}$ peaks at the same applied strain can be pointed to the different definition of the inlet boundary conditions discussed in the 2D setup section. Nevertheless, the graph shows that two-dimensional simulations already confirm the trend of decreasing NO$_{\rm x}$ emissions with increasing strain rate. Similar considerations hold for the trend of NO flux reported in the left y-axis of Figure \ref{fig:NOfluxes}, which is calculated as
\begin{equation}
    \label{eq:NOflux1D}
    \Phi_{\rm NO,1D} = \int_{-L/2}^{L/2} \rho Y_{\rm NO} dx.
\end{equation}
\par
\begin{figure}[t]
	\centering
	\begin{subfigure}{0.4\textwidth}
         \centering
         \includegraphics[width=\textwidth]{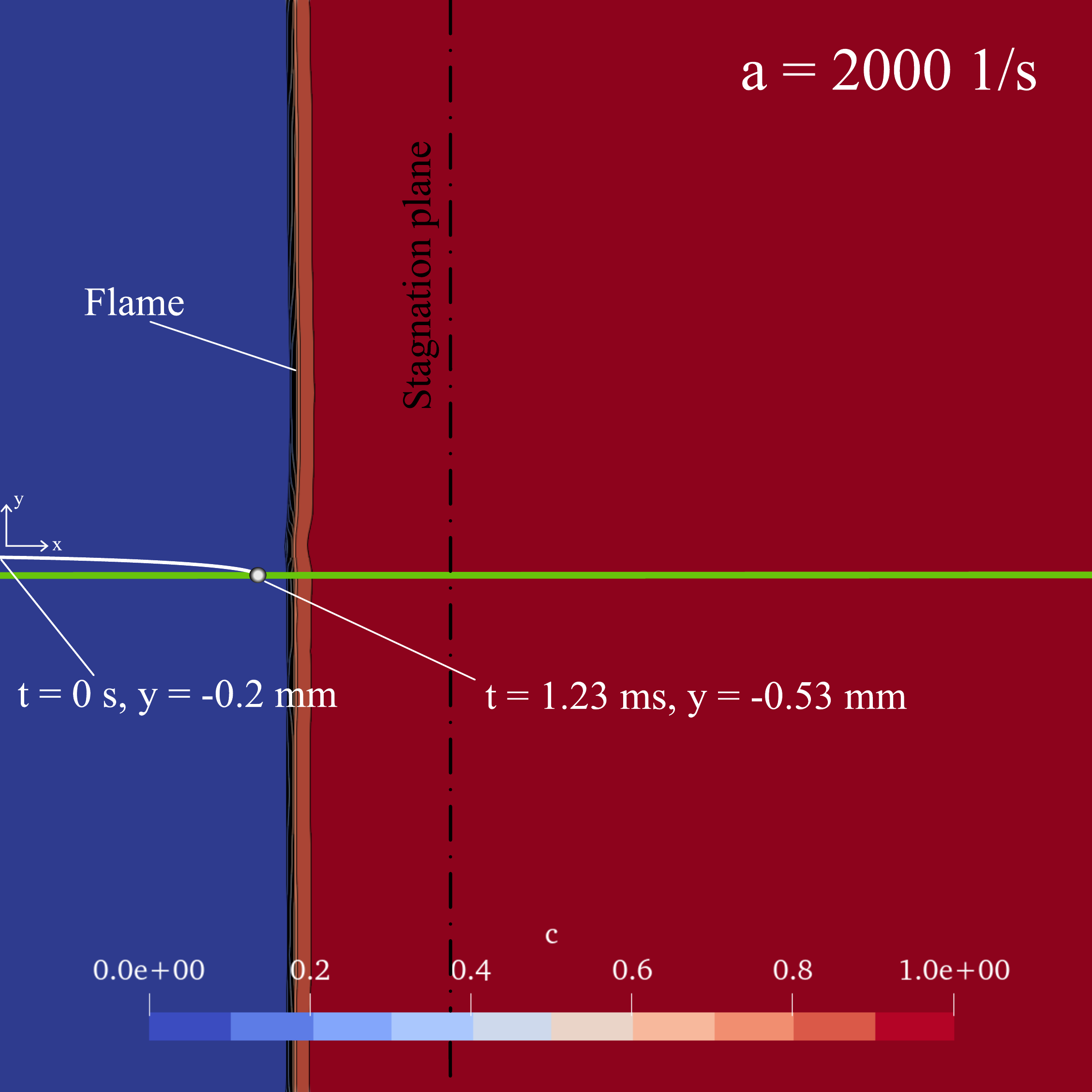}
         \caption{$a=2000 \, {\rm s}^{-1}$.}
         \label{fig:streamline1_S2000}
     \end{subfigure}
     \begin{subfigure}{0.4\textwidth}
         \centering
         \includegraphics[width=\textwidth]{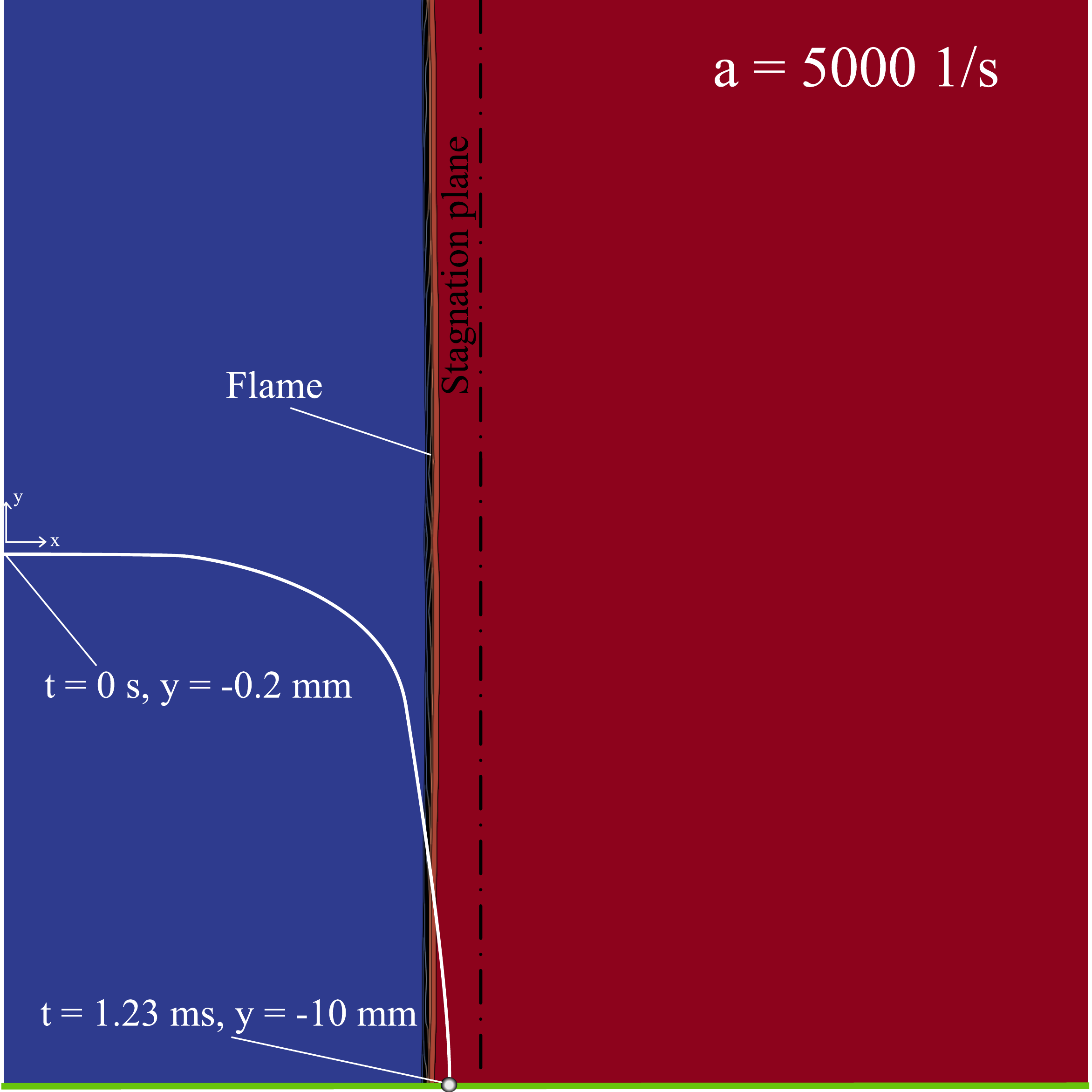}
         \caption{$a=5000 \, {\rm s}^{-1}$.}
         \label{fig:streamline1_S5000}
     \end{subfigure}
	\caption{Travelling path of the a streamline originating at $y=-0.2$ mm on a contour plot of the progress variable $c$ based on $Y_{ \rm H_2O}$.}
	\label{fig:streamline1}
\end{figure}

One might object that NO is a slow species which would form downstream of the flame in the hot products region, and this formation in the products side is prevented by the right boundary condition, where $Y_{\rm NO}(x=L/2)=0$ is imposed. In fact, in the present configuration $Y_{\rm NO}$ peaks at the stagnation plane to then suddenly drop to fulfill the boundary condition, as clearly visible in Figure \ref{fig:NOvsx}. 
Nevertheless, the investigation of the 2D simulations, where the NO is free to form in the $y$-direction (tangential to the flame in Figure~\ref{fig:counterflow_config}), suggests this not to be the case, i.e. NO is suppressed at high strain by some other means. 
Hence, the focus is shifted to the understanding of the emission phenomena in the flame-tangential direction of the two-dimensional simulations. 
One can compute the two-dimensional NO flux for the two setups investigated by integrating on the 2D-domain surface as follows:
\begin{equation}
    \label{eq:NOflux2D}
    \Phi_{\rm NO,2D} = \int_{A} \rho Y_{\rm NO} dA.
\end{equation}
This flux is reported on the right y-axis in Figure \ref{fig:NOfluxes}. It can be immediately observed that the same decreasing NO flux trend is observed for the 1D and 2D cases. Furthermore, from the analysis of the 2D data, it is highlighted that
\begin{equation}
   \Phi_{\rm NO,2D} \approx \Phi_{\rm NO,1D} \cdot L_{\rm 2D},
\end{equation}
i.e. the 1D flux of NO calculated across the center line multiplied by the vertical dimension is approximately equal to the 2D flux. These considerations raise confidence on the fact that the decrease of NO emissions observed in 1D simulations and in the 2D domain centreline as strain is increased is not compensated by any additional NO formation in the vertical direction. This evidence can be further proved following the same streamline at the two different strain rate cases investigated with OpenFOAM. The streamline investigated originates at $y=-0.2$ mm, such that it crosses the flame and thus travels in the products and exits the domain without clashing with the stagnation plane. As shown in Figure \ref{fig:streamline1_S5000}, the particle residence time inside the $a=5000 \, {\rm s}^{-1}$ domain is estimated around $\Delta t=1.23 \, \rm{ms}$. The path that the particle originating in the same position has travelled in the same time interval in the $a=2000 \, {\rm s}^{-1}$ domain is reported in Figure \ref{fig:streamline1_S2000}. Hence, a domain cut is taken at the $y$-location the particle has reached at $\Delta t$ in the two cases  (green line in Figure \ref{fig:streamline1}), and the NO fluxes computed along the line cuts are reported in Table \ref{tab:streamline1}.
\begin{table}[b]
	\centering
	\caption{Flux of NO after the same particle has travelled for $\Delta t = 1.23$ ms at two different applied strain rates.}
	\vspace{8pt}
	\begin{tabular}{c c c}
		\hline\noalign{\smallskip}
		$a$ [$ \, {\rm s}^{-1}$] & $y(\Delta t)$ [mm] & $\Phi(y)$ [g/cm$^2$]  \\
		\noalign{\smallskip}\hline\noalign{\smallskip}
		2000 & -0.53 & 9.51 $\cdot$ 10$^{-10}$ \\
		5000 & -10 & 2.56 $\cdot$ 10$^{-10}$ \\
		\noalign{\smallskip}\hline
	\end{tabular}
	\label{tab:streamline1}
\end{table}
Not only do the data show a still lower NO flux at higher strain rates, but it can be also observed that the values are very close to the ones at the center line (reported in the graph in Figure \ref{fig:NOfluxes}). Therefore, it is further proved that there is no additional NO formation in the vertical direction at higher strain rates. It can be concluded that, at least within the studied framework, the observed NO emissions decrease with increasing strain is not a consequence of the specific counter-flow configuration chosen and of the boundary-preconditioned numerical setup, but is instead a direct effect of the increased tangential velocity gradients on the hydrogen flame combustion and dissociation reactions.

\subsection*{NO formation pathways} \label{sec:NOpathways}
\begin{figure}[t]
	\centering
	\begin{subfigure}{0.5\textwidth}
         \centering
         \includegraphics[width=\textwidth]{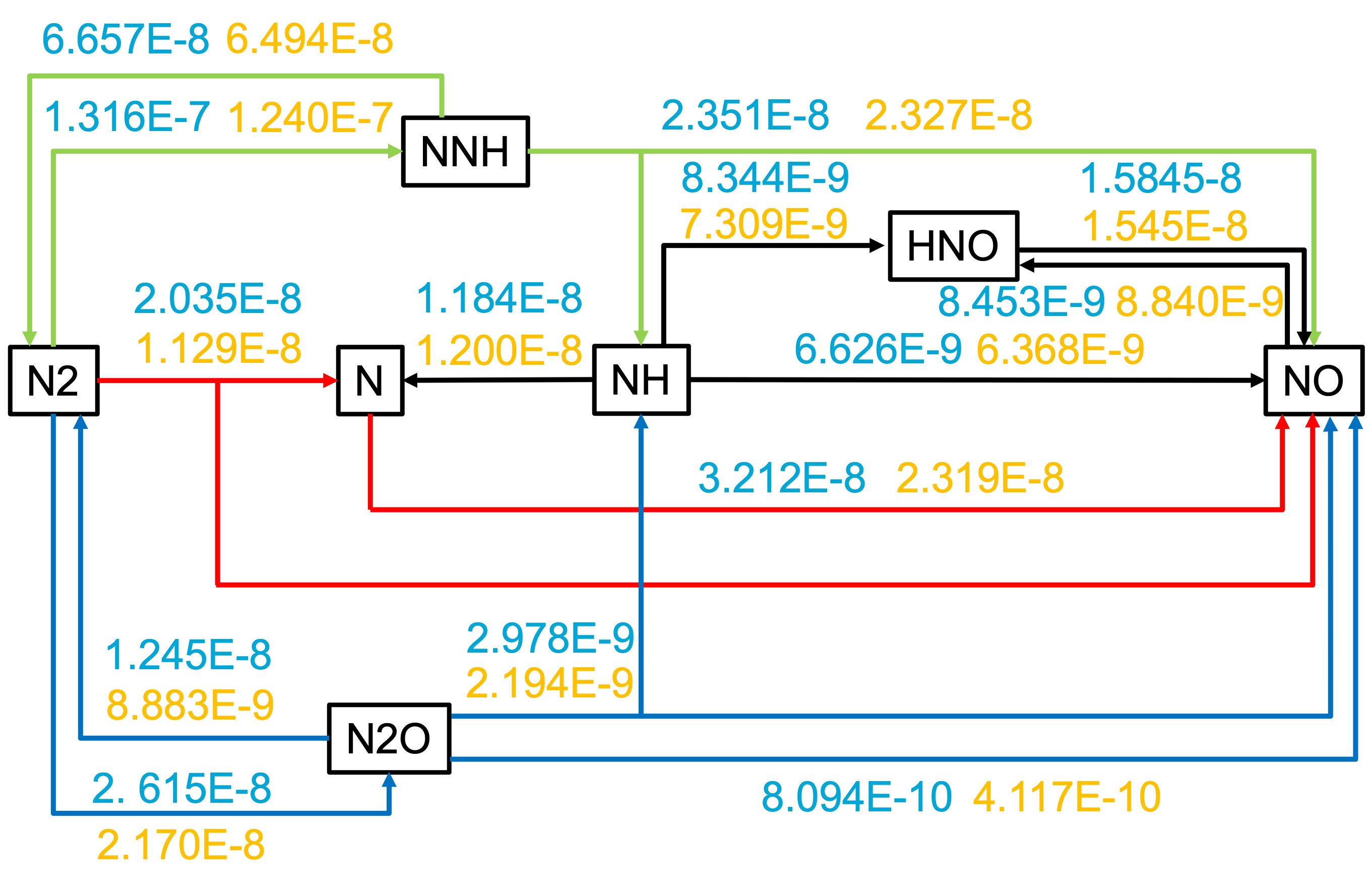}
         \caption{}
         \label{fig:NOpathways}
     \end{subfigure}
     \begin{subfigure}{0.45\textwidth}
         \centering
         \includegraphics[width=\textwidth]{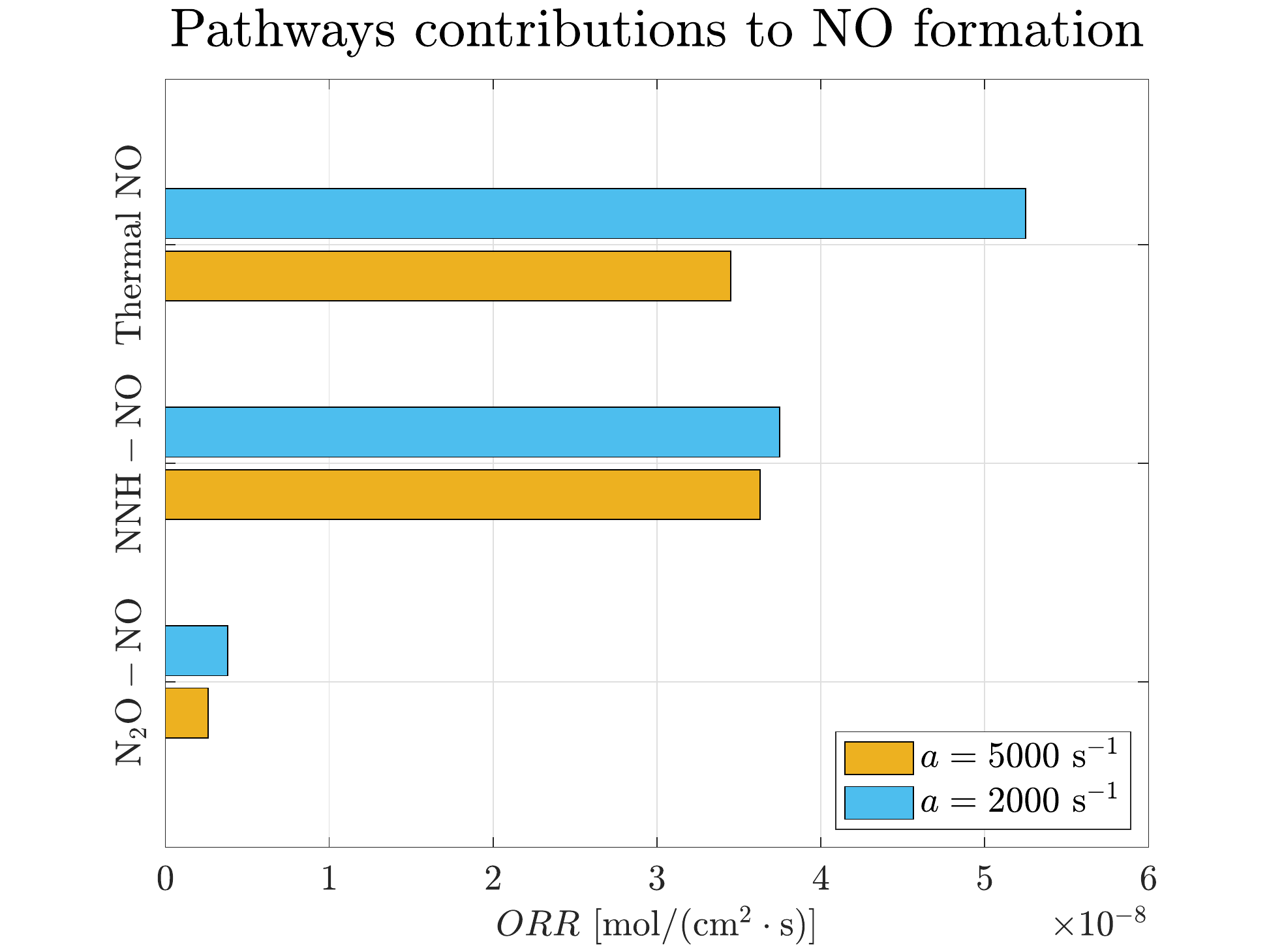}
         \caption{}
         \label{fig:NObar}
     \end{subfigure}
	\caption{(a) Chart diagram showing the overall reaction rates [mol/(cm$^2\cdot$s)] along the NO formation pathways. Thermal NO pathway is highlighted in red, NNH-NO pathway in green, NNH-NO-HNO intermediate pathway in black, and N$_2$O-NO pathway in blue. (b) 
    Contribution of each NO formation pathway to the total NO formation. For strain level $a=2000 \, {\rm s}^{-1}$ and $a=5000 \, {\rm s}^{-1}$.}
	\label{fig:NOanalysis}
\end{figure}

The focus is now shifted to the evaluation of the NO formation pathways, with the goal of further understanding the physical reasons behind the observed NO decrease with strain. As carbon species are not involved in hydrogen combustion, the prompt NO$_{\rm x}$ formation pathway is not considered, and the only three pathways evaluated are thermal NO, NNH-NO and N$_2$O-NO. CHEM1D solutions data are used for the purposes of the present NO formation routes investigations, since the trends of NO emissions with strain predicted by one-dimensional and two-dimensional simulations have been shown to compare with good approximation (see Figure \ref{fig:NOfluxes}). Two high applied strain rates setups of $a=2000 \, {\rm s}^{-1}$ and $a=5000 \, {\rm s}^{-1}$ are analysed. 
For a sample reaction $r$ involved in a pathway $\nu_A A + \nu_B B \leftrightarrow \nu_C C + \nu_D D$, the forward reaction coefficient is obtained with Arrenhius law, and thus the reverse reaction coefficient can be found \cite{poinsot2005theoretical}:
\begin{equation}
    \label{eq:arrhenius}
    K_{f,r}(T)=A_rT^{\beta_r} e^{\left( \frac{-E_{a,r}}{R_0 T} \right) }, \hspace{1cm} K_{r,r}(T)=\frac{K_{f,r}}{\left(\frac{p_a}{R_0 T}\right)^{\sum_s \nu_s} e^{\left(\frac{\Delta S_r^0}{R_0}-\frac{\Delta H_r^0}{R_0 T}\right)}},
\end{equation}
where $A_r$, $\beta_r$ and $E_{a,r}$ are found in the \texttt{GRI3.0} mechanism reaction web page \cite{GRI30}, and $\Delta S_r^0$ and $\Delta H_r^0$ are respectively the reaction enthropy and enthalpy that can be found as a function of temperature with JANAF polynomials. Thus, the forward and reverse reaction rates are found as a function of $T(x)$ introducing respectively the reactants and products concentrations:
\begin{equation}
    \label{eq:omega}
    \dot{\omega}_{f,r} = K_{f,r}(x) \cdot [A(x)]^{\nu_A} [B(x)]^{\nu_B}, \hspace{1cm} \dot{\omega}_{r,r} = K_{r,r}(x) \cdot [C(x)]^{\nu_C} [D(x)]^{\nu_D}.
\end{equation}
Three-body and pressure-dependent reactions have been treated with dedicated formulas \cite{kee1989chemkin}.
Hence, following the formulation of \citet{ning2017effects}, the overall reaction rate ($ORR$) of the reaction $r$ is obtained by integrating the reaction rate across the longitudinal (flame-normal) direction:
\begin{equation}
    \label{eq:ORR}
    ORR_r = \int_{-L/2}^{L/2} \dot{\omega}_r(x) dx.
\end{equation}
Summing the contribution of the single reactions $ORR$ along a specific route, a chart diagram showing the NO formation pathways at $a=2000 \, {\rm s}^{-1}$ and $a=5000 \, {\rm s}^{-1}$ is obtained and is reported in Figure \ref{fig:NOpathways}. While all the reaction rates involved in the pathways show a decrease at higher strain, it can already be observed that this decrease is remarkably more consistent considering the reactions involved in the thermal NO pathway.
Hence, the $ORR$ of the reactions directly producing NO in each pathway are summed and the bar chart in Figure \ref{fig:NObar} is obtained. The histogram shows that the main NO production pathways involved in lean hydrogen flames at high strain rates are both the thermal and the NNH. Considering syngas fuel mixtures and twin counter-flow configuration, \citet{ning2017effects} have shown with a similar analysis that the NNH pathway was by itself predominant, even with a high hydrogen content. The reason of the discrepancy with the present study is probably related to the lower equivalence ratio of their investigation ($\phi=0.5$), which may decrease the maximum flame temperature and thus the weight of thermal NO pathway on the total NO formation process. In addition, they conclude that with a high hydrogen content in the fuel and high strain rates the NO production is exacerbated. Once again, the fuel mixture and the equivalence ratio, along with the maximum strain rates achieved and the counter-flow configuration chosen do not allow for a direct comparison even in the trends with the present study, which instead investigates on completely novel setup with 100\% hydrogen fuel and very high strain rates. \\
Figure \ref{fig:NObar} confirms that the main decrease of NO production at higher strain rates is associated to the thermal pathway. \citet{speth2009using} have shown that for syngas with a high content of H$_2$, the flame temperature has an increasing trend with strain within their limited strain range, i.e. up to $a=500 \, {\rm s}^{-1}$. Furthermore, extending the strain range would probably result in a non-monotonic function, as their study shows for fuel mixtures with less hydrogen content. NO$_{\rm x}$ emissions can be expected to follow directly the flame temperature function. In contrast, the present study shows that NO$_{\rm x}$ emissions decrease monotonically in a very broad range of strain rates, thus indicating that the emissions function does not follow directly the temperature function. Hence, the reason behind the observed NO decreasing trend may be not only imputable to a change in the flame temperature.
This consideration suggests that a more extensive analysis should be performed on the local temperature and concentration of the radicals involved in the pathway reactions in order to further shed light on the physical explanation behind the phenomenon of decreasing NO$_{\rm x}$  with strain in lean premixed hydrogen flames.

\section*{Conclusion}
Detailed-chemistry one-dimensional and two-dimensional DNS analyses are conducted on pure hydrogen lean premixed and strained flamelets in a counter-flow reactants-to-products configuration. A broad range of strain rates is investigated, up to very high levels which have been rarely considered before in the literature, and which hydrogen is shown to be able to sustain without experiencing extinctions. Both 1D and 2D simulations show that NO$_{\rm x}$ emissions have a decreasing trend as the strain rate is increased. This is shown for the first time for pure hydrogen fuel and lean flamelets. The analysis of the 2D results also indicates that there is no compensating NO$_{\rm x}$ formation in the second dimension, and thus that the trend observed is not a consequence of any possible setup preconditioning. Therefore, the greater tangential velocity gradients may have a direct effect on the flame, somehow affecting combustion efficiency and consequently the rate of reactions through which NO$_{\rm x}$ is formed. \\
In-depth analyses of the NO$_{\rm x}$ formation mechanisms and reaction rates further underline that the main contribution to the decreasing trend arises from the thermal NOx pathway. Future work will focus on a more extensive local temperature and radical concentration study on the setup investigated and at further conditions in order to achieve a deeper understanding of the physics behind the phenomenon. Consumption speed trend with strain and the role of differential diffusion effects are also hydrogen-peculiar features that may play a role and that should be further explored. 
If the emission trend with strain are confirmed in turbulent conditions, the low NO$_{\rm x}$ outcomes of lean premixed and highly-strained hydrogen flames can be exploited in future technological applications of hydrogen combustion systems.

\section*{Acknowledgments}
AP and IL acknowledge the Dutch Ministry of Education and Science for providing funding support to this project via the Sector Plan scheme. BK and IL further acknowledge support from the Engineering and Physical Sciences Research Council, grant n. EP/T028084/1.

\bibliographystyle{unsrtnat}
\bibliography{references}

\end{document}